\documentclass{article}
\usepackage{arxiv}

\usepackage[utf8]{inputenc} 
\usepackage[T1]{fontenc}    
\usepackage{hyperref}       
\usepackage{url}            
\usepackage{booktabs}       
\usepackage{amsfonts}       
\usepackage{nicefrac}       
\usepackage{microtype}      
\usepackage{lipsum}		
\usepackage{graphicx}
\usepackage{doi}
\usepackage{subcaption}
\usepackage[numbers]{natbib}

\title{Gaining a better understanding of online polarization by approaching it as a dynamic process}

\author{ \href{https://orcid.org/0000-0003-1634-8856}{Célina TREUILLIER}\\
	Université de Lorraine, CNRS, LORIA\\
	Nancy, FRANCE\\
	\texttt{celina.treuillier@loria.fr} \\
	\And
	\href{https://orcid.org/0000-0003-4252-4526}{Sylvain CASTAGNOS} \\
	Université de Lorraine, CNRS, LORIA\\
	Nancy, FRANCE\\
	\texttt{sylvain.castagnos@loria.fr} \\
 	\And
  	\href{https://orcid.org/0000-0002-7693-1325}{Christèle LAGIER} \\
	Avignon Université, JPEG UPR 3788\\
	Avignon, FRANCE\\
	\texttt{christele.lagier@univ-avignon.fr} \\
 	\And
	\href{https://orcid.org/0000-0002-9876-6906}{Armelle BRUN} \\
	Université de Lorraine, CNRS, LORIA\\
	Nancy, FRANCE\\
	\texttt{armelle.brun@loria.fr} \\
}

\date{}

\renewcommand{\headeright}{}
\renewcommand{\undertitle}{}

\begin{document}
\maketitle

\begin{abstract}
Polarization is often a cliché, its conceptualization remains approximate and no consensus has been reached so far. Often simply seen as an inevitable result of the use of social networks, polarization nevertheless remains a complex social phenomenon that must be placed in a wider context.  To contribute to a better understanding of polarization, we approach it as an evolving process, drawing on a dual expertise in political and data sciences. We compare the polarization process between one mature debate (COVID-19 vaccine) and one emerging debate (Ukraine conflict) at the time of data collection. Both debates are studied on Twitter users, a highly politicized population, and on the French population to provide key elements beyond the traditional US context. This unprecedented analysis confirms that polarization varies over time, through a succession of specific periods, whose existence and duration depend on the maturity of the debate. Importantly, we highlight that polarization is paced by context-related events. Bearing this in mind, we pave the way for a new generation of personalized depolarization strategies, adapted to the context and maturity of debates.
\end{abstract}

\section{Introduction}
\label{sec:intro}

Surveys on polarization proliferate, but no consensus has been reached on the definition of this highly democratic topic \cite{kubin2021role}. The term \textit{polarization} is employed in favor of very heterogeneous analyses \cite{sunstein1999law}. It is becoming the root of all evil. It is for example used to describe the polarization of juries, or the vote in favor of Brexit, as well as for the decisions made in deliberative assemblies or journalistic commentary in the face of the invasion of the US capitol in January 2021. This predigested approach of polarization obscures the fact that these events are also embedded into social phenomena that political science has long been studying. These include the unequal distribution of political competence \cite{carpini1996americans}, resentment \cite{cramer2016politics,hochschild2018strangers} and distrust of political leaders, the shortcomings of political representation \cite{saward2006representative}, contextual effects \cite{huckfeldt1986politics,huckfeldt1995citizens}, the weight of primary groups \cite{lazarsfeld1968people,zuckerman2005social} or of discussion \cite{pattie2009conversation}, and politics avoidance \cite{eliasoph1998avoiding} in the formation of opinions. Polarization is a source of even greater interest as the availability of digital data in virtually unlimited quantities, notably with the advent of social networks, becomes more widespread. In this new informational context, many works attempt to model \cite{sirbu2017opinion,baumann2020modeling,chen2019modeling}, measure \cite{conover2011political, guerra2013measure, becatti2019extracting, cicchini2022news}, and identify \cite{garimella2021political} polarization. These works show that the very format of social networks contributes to exacerbating polarization – that is, in part, artificially co-constructing it – through anonymity and increased selective exposure. Yet these works tend to adopt a purely data-oriented analysis and barely consider the underlying social phenomena, although they are aware of their existence \cite{cinus2022effect}. Finally, by a circular effect, the certainty of the existence of polarization is reinforced by analyses produced on the basis of models that struggle to go beyond a binary identification of polarization (polarized \textit{vs.} non-polarized).

In our view, the main pitfall of these works lies in the fact that polarization is taken for granted in a context of high social and political tension that characterize our society \cite{geschke2019triple}. Two presuppositions, often poorly explained, underlie these approaches. On the one hand, polarization is mainly seen as an effect of social networks \cite{jost2022cognitive, valensise2023drivers}. On the other hand polarization, often approached in terms of affective polarization \cite{kubin2021role}, is viewed as a state that can be merely measured at a given time. The goal of our work is not to question the impact of social media on polarization \cite{kubin2021role,prior2013media}, particularly related to the algorithmic filtering of information that tends to confirm users in their beliefs \cite{pariser2011filter}. Combining contributions from political science and data science, it proposes to draw on the distinction between affective and ideological polarization \cite{kubin2021role} in a temporal perspective. We assume that polarization mechanisms (notably social influence and the role played by persuasive arguments \cite{sunstein1999law}) can be cumulative or successive when polarization processes are observed over the medium to long term and we consider polarization as a politicization process. To validate this hypothesis, we focus on the French context, little studied in the literature, where left-right referents still strongly structure the political and media field. Indeed, while this distinction tends to become blurred for part of the population \cite{marchand2011silent}, but also under the influence of  monopolistic media organization, it still makes sense in a media system inherited from a high level of political homology \cite{hallin2004comparing}.

The Twitter users population is particularly interesting, as it is more aware than the average citizen about political matters \cite{boyadjian2014and,walker2021news}. To this extent, we can suppose that the population of study has a strong level of political interest and we can identify their ideological inclination \cite{barbera2020social}. Besides, discussions on Twitter primarily reflect the concerns and topics addressed by mainstream media. In some respects, this social network appears to be tightly correlated with media framing \cite{russell2014dynamics,mccombs1972agenda}.  In this work, we do not consider that social networks provide a real vision of what is happening among ordinary citizens \cite{prior2013media}. This work rather contributes to overcome the idea that polarization is mainly related to the less informed or misinformed categories of the population. Polarized people are not just the victims of fake news or moral contagion \cite{goldstein1984moral} phenomena that crop up in many analyses of the dynamics of opinion \cite{sirbu2017opinion}. Within this Twitter population, we draw on seminal work about the powerful effect of the environment and selective media exposure of opinion leaders \cite{katz2017personal}. On the overall population, people who polarize are first and foremost those who have the ability to have a constructed opinion on societal debates, and the more asserted this opinion, the less likely it is to be modified \cite{katz2017personal}. 

When polarization is seen as a state, it corresponds to the crystallization of opinions, which has very little chance of being modified \cite{norris2003preaching}. Attempts to depolarize mainly rely on the increase of the diversity of the information recommended to users, to make them confronted with a variety of topics and viewpoints. However, these analyses are not conclusive \cite{helberger2018diversity, heitz2022benefits, joris2019divnrs} as diversity may even further reinforce polarization among users who become resistant to the confrontation of ideas too far from their own \cite{bail2018opposing}. Furthermore, modeling polarization at a given point in time may be completely obsolete sometime later, since it does not consider opinion dynamics. These limits in mind, some works propose to study the evolution of the overall polarization over time \cite{boyadjian2014and,russell2014dynamics,mccombs1972agenda}. However, to the best of our knowledge, most of these works have a global approach, \textit{i.e.} study of polarization as a whole,  and few of them address the temporal evolution of associated polarization behaviors. That is why we propose to sequence polarization in order to better analyze it as a politicization process. 

Although multiple topics and societal issues have been of interest in the literature, we can see that they are all mature topics, \textit{i.e.} topics that have been discussed for months, even years, about which people have had time to take a stand and express clear opinions \cite{garimella2018echo}. As a consequence, the question of how people behave when a new controversial topic emerges remains understudied. An in-depth analysis of the dynamics of polarization about such new controversial topics will help understand how people polarize in the early stages of an emerging debate. 

Concretely, we study temporal polarization behaviors on Twitter about the Ukraine conflict that strongly intensified from February 24, 2022, the date on which the Russian army invaded Ukraine, marking the start of the conflict. The period of study chosen is between January and July 2022. Although this conflict arose from 2014 with the annexation  of Crimea, in 2021 this topic was no longer discussed in the French media for several years.  We also analyze a more mature debate, about which users should already have taken a position about it. We chose the COVID-19 vaccine debate, with the same period of study. For both debates, we distinguish  between users belonging to each  community (pro-vaccine \textit{vs.} anti-vaccine, pro-Ukraine \textit{vs.} pro-Russia). 

This work is one of the first works that addresses polarization from a sequential perspective, and highlights important elements about the evolution of polarization, the impact of the maturity of the debate, and the influence of the context. We confirm that user polarization not only evolves through time, but also differently according to the maturity of the topic. This evolution is not erratic, and specific periods of polarization are identified. More importantly, patterns of such periods are common to debates. They confirm that polarization is a process in which users tend to gradually and naturally come close to extremes and polarize. The evolution of the polarization process can be disturbed by context-related events (covered by the media and discussed on Twitter), which provoke a reset in the pattern of periods and foster users' interactions with the opposing community. The duration of this depends on the maturity of the debate and the nature of the event. We thus assume that cycles of polarization occur, cadenced by the context such as the news events. This opens up opportunities for depolarization strategies that are not only personalized, but also related to the appearance of specific news events. This goes against current depolarization strategies that simply consider diversity.

\section{Results}
\label{sec:results}

\subsection{An aggregate analysis of polarization}
\label{subsec:aggregate}

We first adopt an aggregate analysis of polarization and compare polarization within and between debates: the COVID-19 vaccine and the Ukraine conflict debates. The aggregate analysis relies on an automatic clustering of users, based on their retweet activity (interactions). We consider that clusters formed are made up of users adopting similar polarization behaviors.
We exploit three factors. One opinion factor, computed from users’ interactions with each of the confronting communities, represents the diversity of opinions (in the Ukraine conflict debate for example, 1 represents an extreme polarization in the pro-Ukraine community and 0 an extreme polarization in the pro-Russia community, while 0.5 represents a fair distribution between both communities). Two source factors, computed from users’ interactions on sources of information, represent to which extent a user accesses a large set of sources in either community (with 1 representing an access to a unique source and 0 the access to all sources of information in the community). Identified clusters thus capture the polarization of users according to both their degree of belonging to one or other of the communities (opinion) and the diversity of sources they interact within each of these communities (sources). 

\begin{figure}[h!]
\centering
\begin{subfigure}[t]{0.45\textwidth}
  \includegraphics[width=1\linewidth]{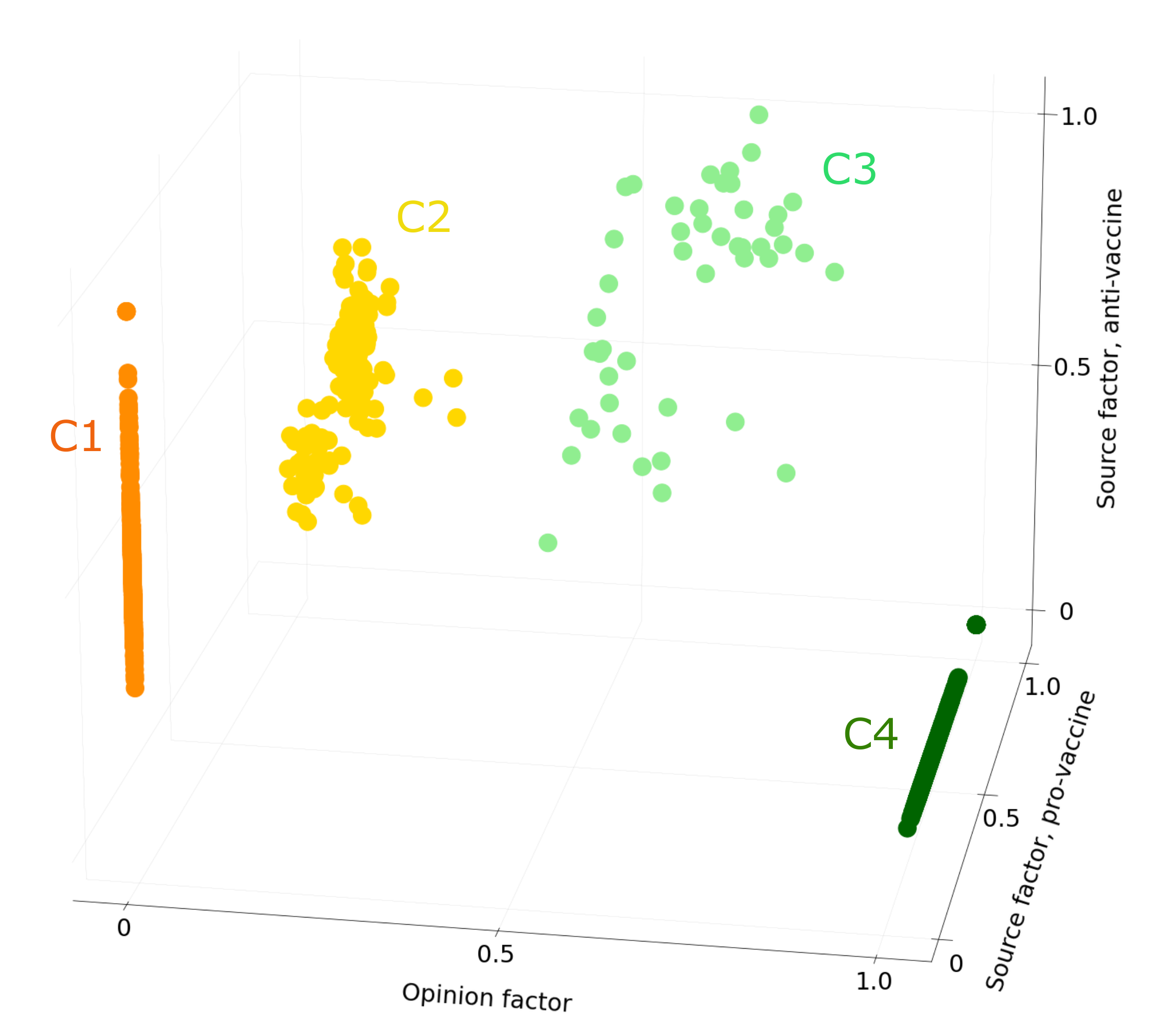}
  \caption{COVID-19 vaccine debate.}
  \label{fig:clusters_aggregate_vaccine}
\end{subfigure}
\begin{subfigure}[t]{0.45\textwidth}
    \centering
    \includegraphics[width=1\linewidth]{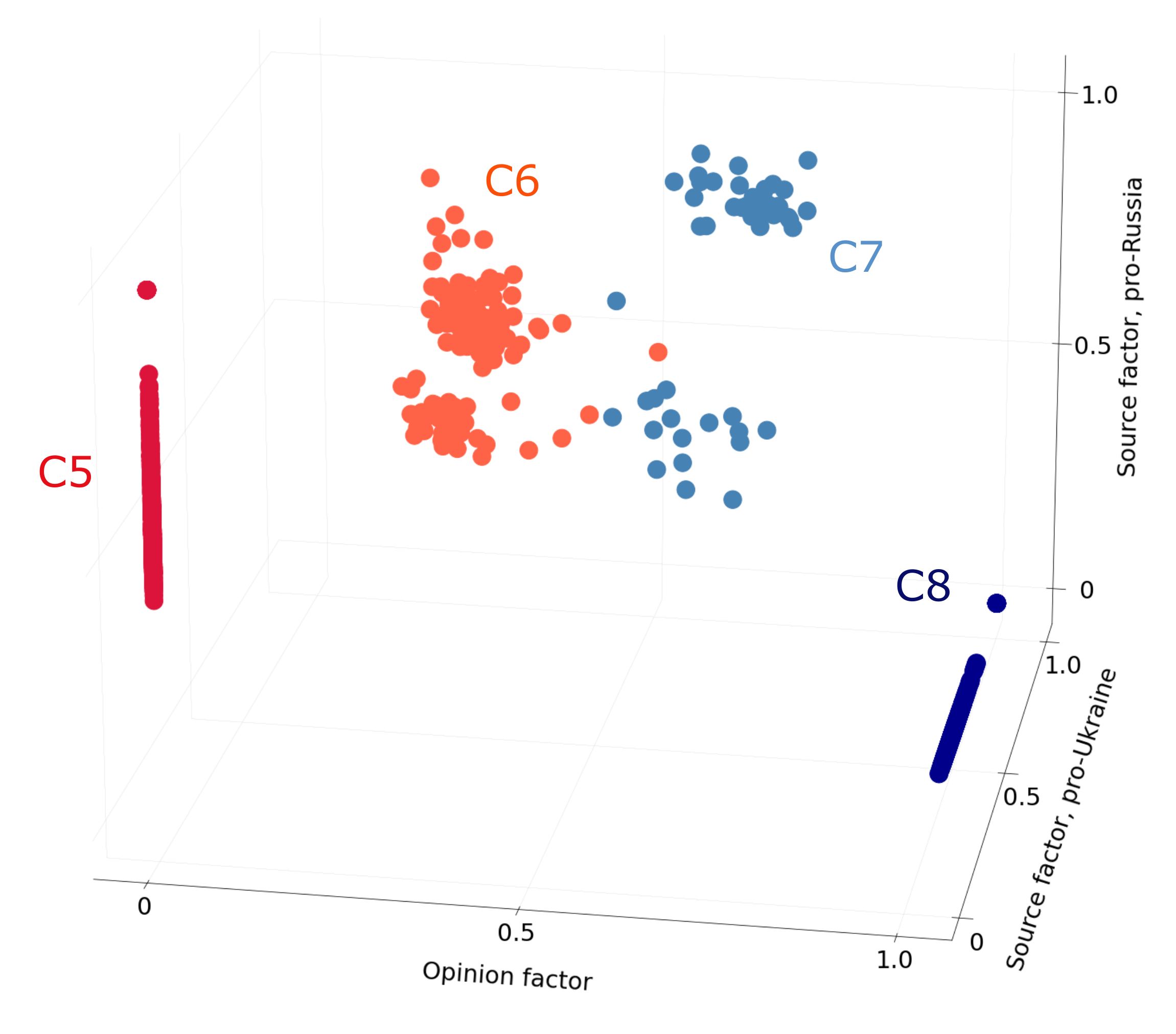}
    \caption{Ukraine conflict debate.}
    \label{fig:clusters_aggregate_ukraine}
\end{subfigure}
\caption{\textbf{Clusters resulting from the aggregate analysis.} Figure (a) presents the clusters identified among users interacting about the COVID-19 vaccine debate (n=1000). The proportions of users in each cluster are as follows: C1 = 36\%, C2 = 14\%, C3 = 5\%, and C4 = 45\%. Figure (b) presents the clusters identified among users interacting about the Ukraine conflict debate (n=1000). The proportions of users in each cluster are as follows: C5 = 34\%, C6 = 16\%, C7 = 5\%, and C8 = 45\%.}
\label{fig:clusters_aggregate}
\end{figure}

For each debate, the optimal number of clusters is 4. The clustering is of high quality as Silhouette indexes are equal to 0.85 and 0.87 (the higher the better) and Davies-Bouldin indexes equal to 0.35 and 0.32 (the lower the better) for the COVID-19 vaccine and Ukraine conflict debates respectively. Identified clusters are presented in a three-dimensional space, where each dimension corresponds to one factor (opinion factor and the two source factors) (See Figure \ref{fig:clusters_aggregate}).

In each debate findings are similar. We identify two major clusters corresponding to users who interact with one community only (C1 and C4, C5 and C8), whatever is the diversity of sources they access in this community. We refer to these users as polarized users. The two other clusters  (C2 and C3, C6 and C7) are made up of users interacting with both communities,  but who still have a preference for one side of the debate, with which they are more active. In their favored community, users interact with multiple sources. In the other community, they interact with a varying number of sources, ranging from a unique to all sources. These users thus confront opposing viewpoints. We refer to them as intermediate users. They represent a significant part of the population: 19\% for the COVID-19 vaccine debate and 21\% for the Ukraine conflict debate. Such a significant representation was not expected given the polarizing nature of the debates. We also note that these proportions are similar for both debates. We might have thought that the number of intermediate users involved in the emerging debate would have been higher. 

To summarize, this aggregate analysis, based on an individual and multi-factorial representation of polarization, contributes to differentiating between four distinct polarization classes. It therefore goes beyond the simple distinction between polarized and non-polarized users from the literature \cite{becatti2019extracting, treuillier23}. In particular, it contributes to highlighting the existence of clusters of intermediate users, with a moderate level of polarization. Although the two selected debates are not related and have different maturity levels at the time of data collection, the clusters formed are similar in number and interpretation, and are made up of a similar proportion of users. In light of this aggregate analysis, one could conclude that the strength of polarization does not seem to depend on the maturity of the debate. Nevertheless, we wonder how these clusters evolve over time, both in terms of number and nature, and if this evolution differs according to the maturity of the debate.

\subsection{A time-aware analysis of polarization}
\label{subsec:time-aware}

\begin{figure}[h!]
    \centering
    \includegraphics[width=1\linewidth]{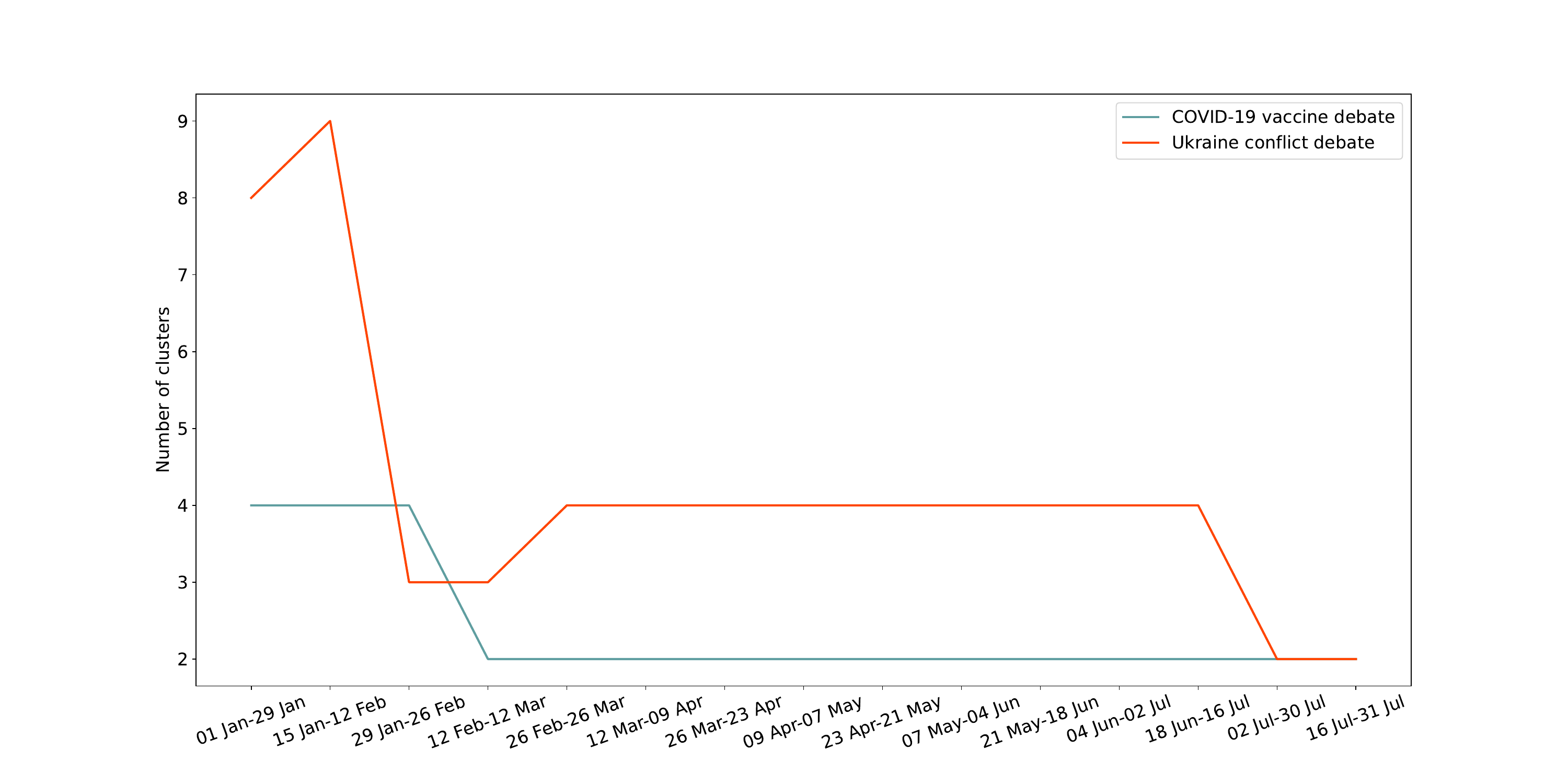}
    \caption{\textbf{Evolution of the number of identified clusters.} The figure shows the evolution of the number of identified clusters over each timeframe.}
    \label{fig:evolution_clusters}
\end{figure}

Recall that in this work we approach polarization as an evolving process, by which it may be possible to identify if the clusters previously identified evolve over time.  The  study relies on the analysis of clusters of users  automatically identified within timeframes. Each timeframe is 4 weeks long and consecutive timeframes have a 2-week overlap. 
By looking at the number of identified clusters in each timeframe, we can observe that it actually varies through time (See Figure \ref{fig:evolution_clusters}). This shows that user polarization actually evolves, whatever is the maturity of the debate. While variations were conceivable for the Ukraine conflict debate which was recent at the time of data collection, such variations are particularly intriguing for the COVID-19 debate, which had been discussed for a long time.

\begin{figure}[h!]
\centering
\begin{subfigure}[t]{0.45\textwidth}
  \includegraphics[width=1\linewidth]{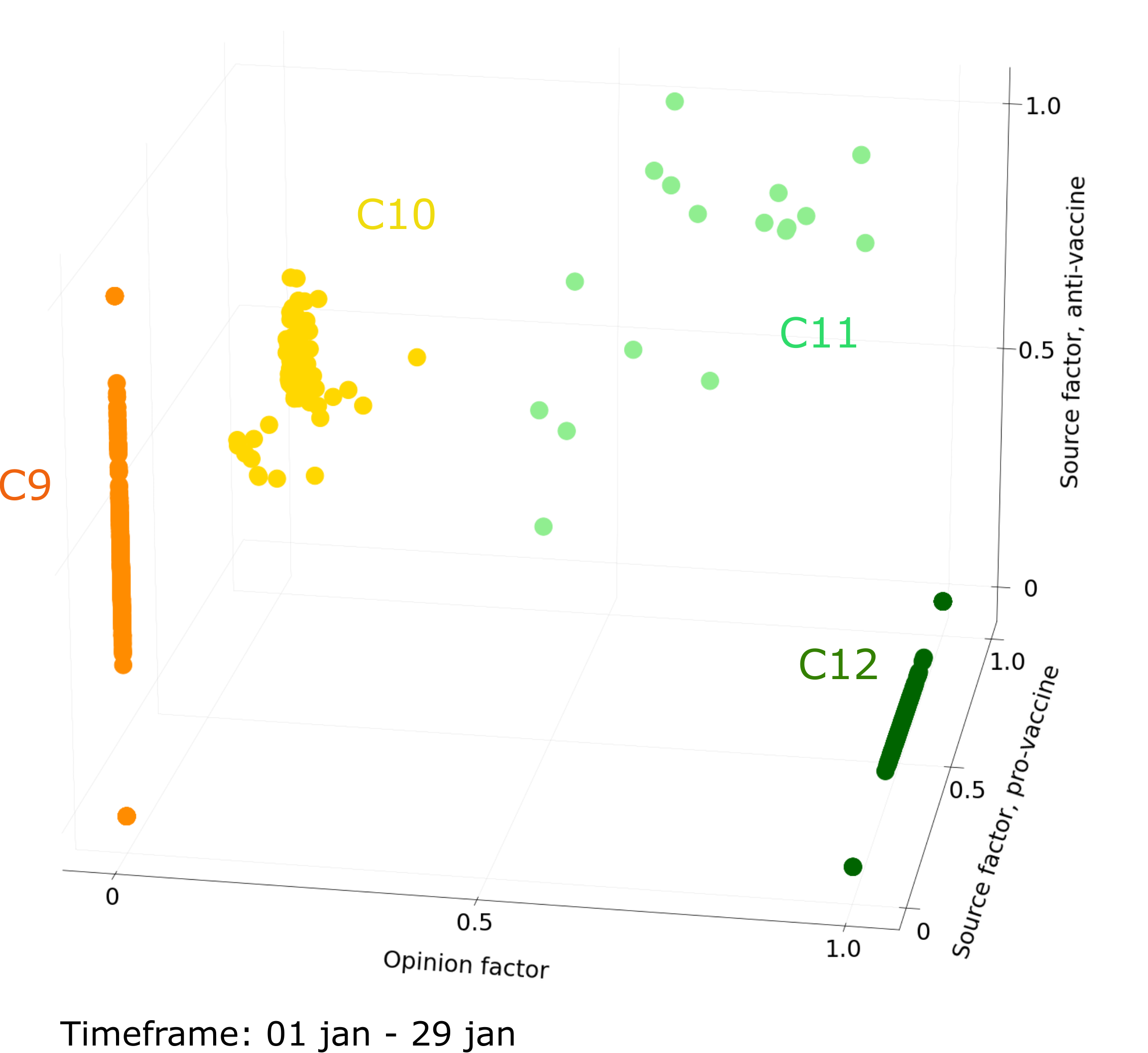}
  \caption{4 clusters}
  \label{fig:4clusters_vaccine}
\end{subfigure}
\begin{subfigure}[t]{0.45\textwidth}
  \includegraphics[width=1\linewidth]{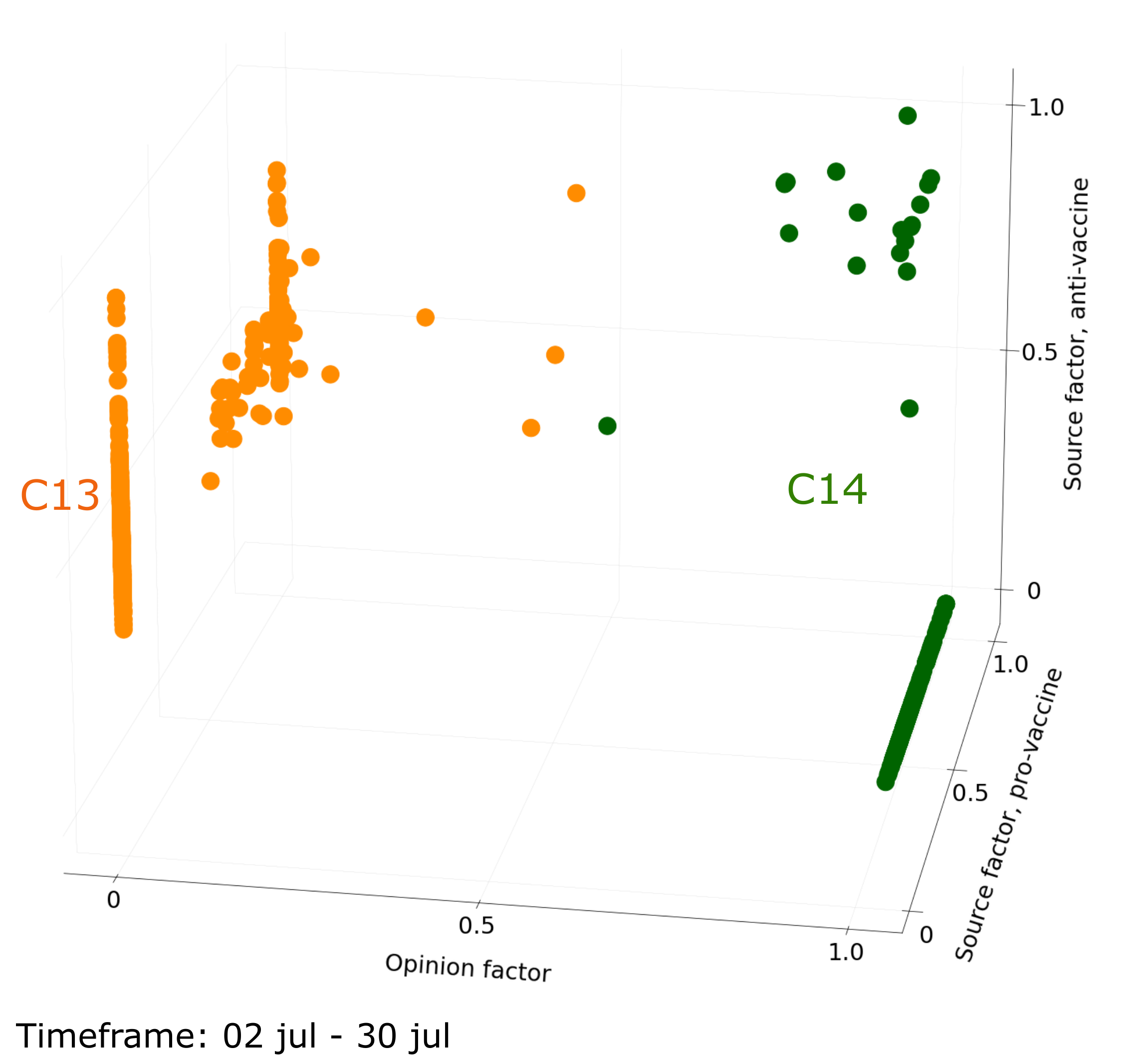}
  \caption{2 clusters}
  \label{fig:2clusters_vaccine}
\end{subfigure}
\caption{\textbf{Clusters of users interacting about the COVID-19 vaccine debate (n=685 users active on at least 80\% of timeframes).} Figure (a) presents the clusters identified among users during a 4-clusters timeframe, extending from January 1 to January 29, 2022. The proportions of users in each cluster are as follows: C9 = 53\%, C10 = 13\%, C11 = 2\%, and C12 = 32\%. Figure (b) presents the clusters identified among users during a 2-clusters timeframe, extending from July 2 to July 31, 2022. The proportions of users in each cluster are as follows: C13 = 66\%, C13 = 34\%}
\label{fig:clusters_vaccine}
\end{figure}

In addition, this evolution differs between debates. For the long-lasting COVID-19 vaccine debate, the number of clusters varies between 2 and 4 clusters. During the three first timeframes, ranging from January 1 to February 26, 4 clusters are identified. These clusters (C9, C10, C11, and C12 in Figure \ref{fig:4clusters_vaccine}) are consistent with those identified in the aggregate analysis of polarization (C1, C2, C3, and C4 in Figure \ref{fig:clusters_aggregate_vaccine}). Indeed, two clusters (C8 and C12) are made up of polarized users, interacting in a unique community, while two clusters (C10 and C11) are made up of intermediate users, interacting in both communities,  with unbalanced interactions with one community. From the 4th timeframe till the end of the dataset timespan, only 2 clusters are discriminated (C13 and C14 in Figure \ref{fig:2clusters_vaccine}). During these timeframes, all users are identified as polarized. 

\begin{figure}[h!]
\centering
\begin{subfigure}[t]{0.32\textwidth}
  \includegraphics[width=1\linewidth]{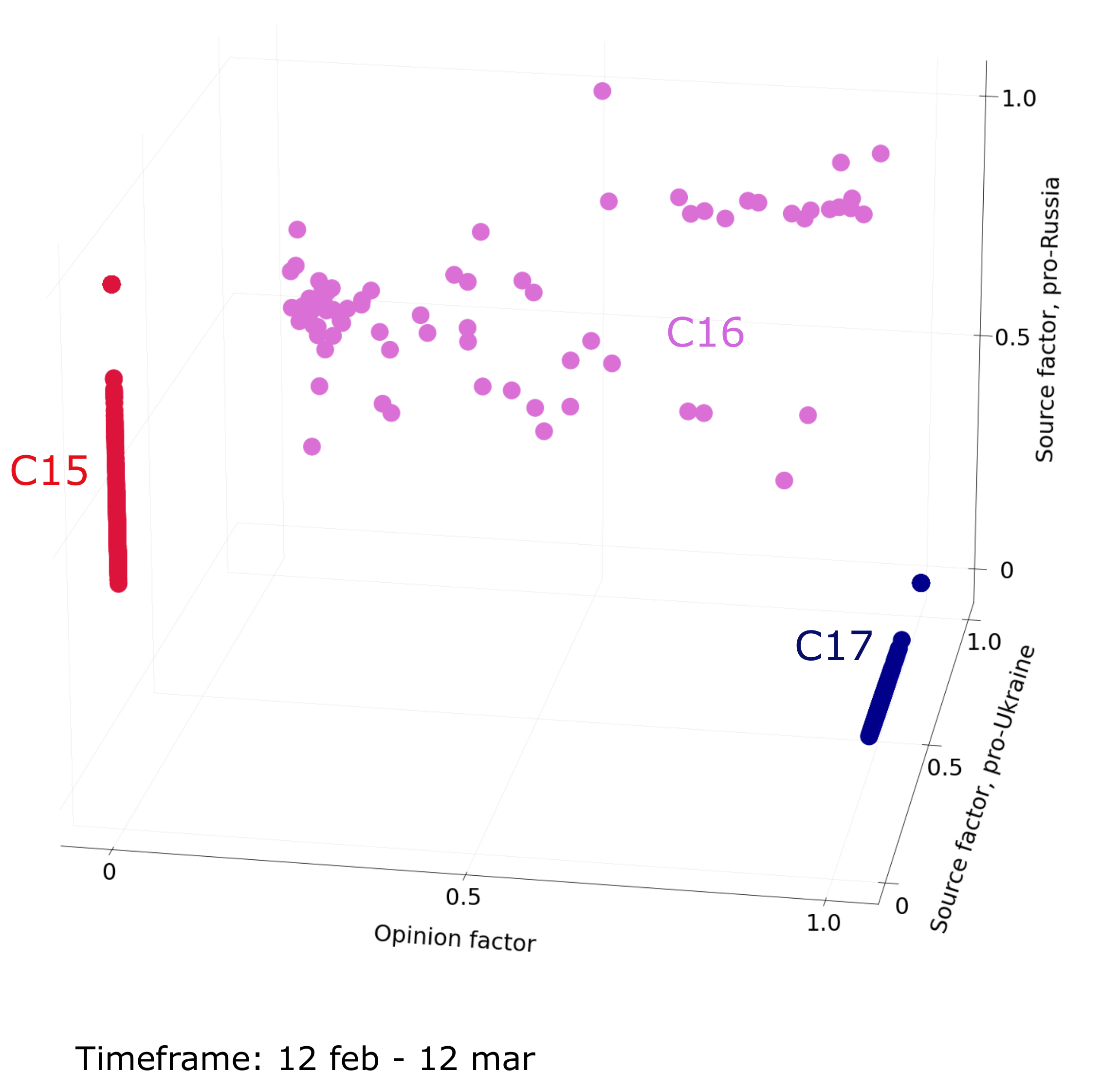}
  \caption{3 clusters}
  \label{fig:3clusters_ukraine}
\end{subfigure}
\begin{subfigure}[t]{0.32\textwidth}
  \includegraphics[width=1\linewidth]{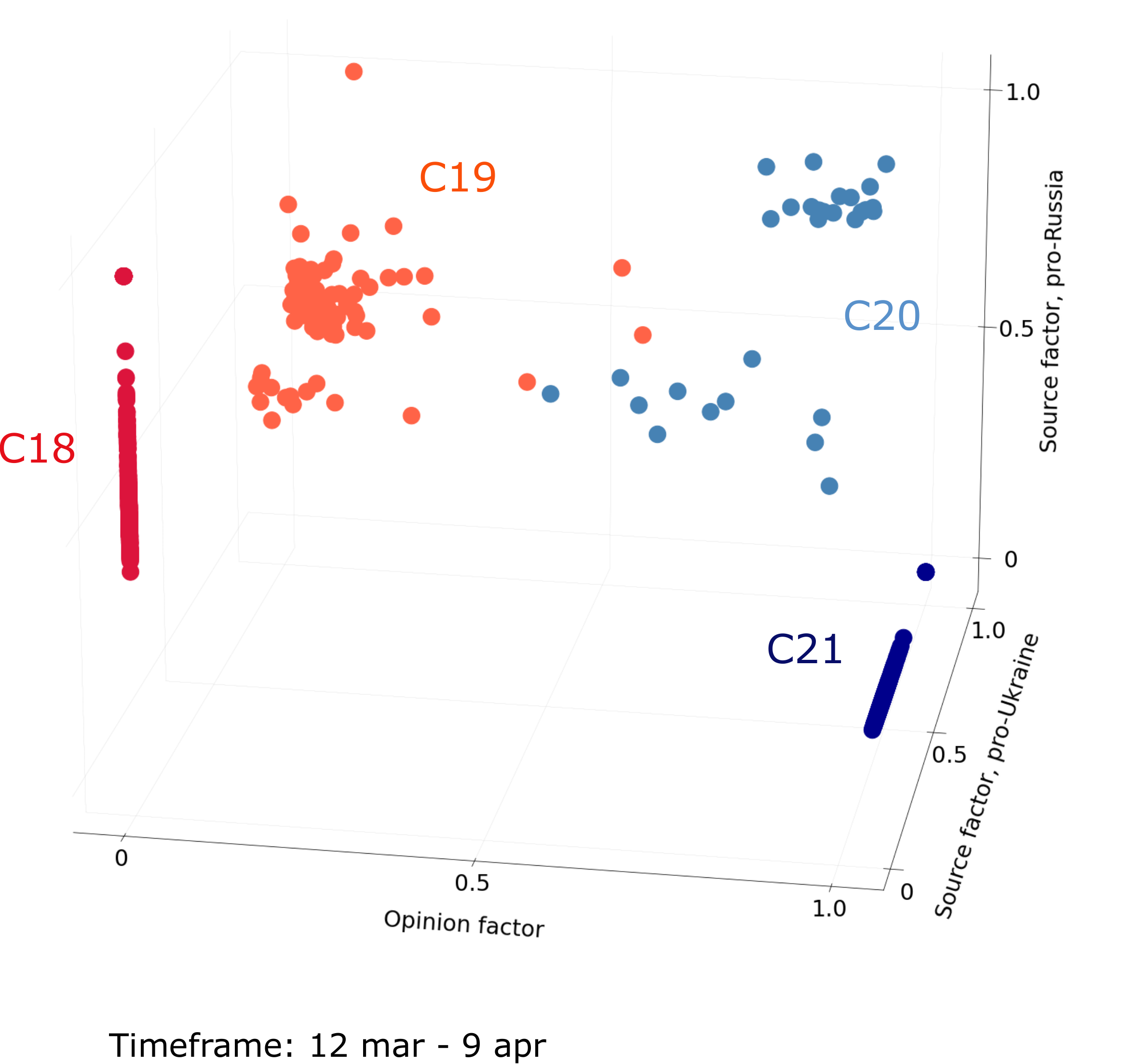}
  \caption{4 clusters}
  \label{fig:4clusters_ukraine}
\end{subfigure}
\begin{subfigure}[t]{0.32\textwidth}
  \includegraphics[width=1\linewidth]{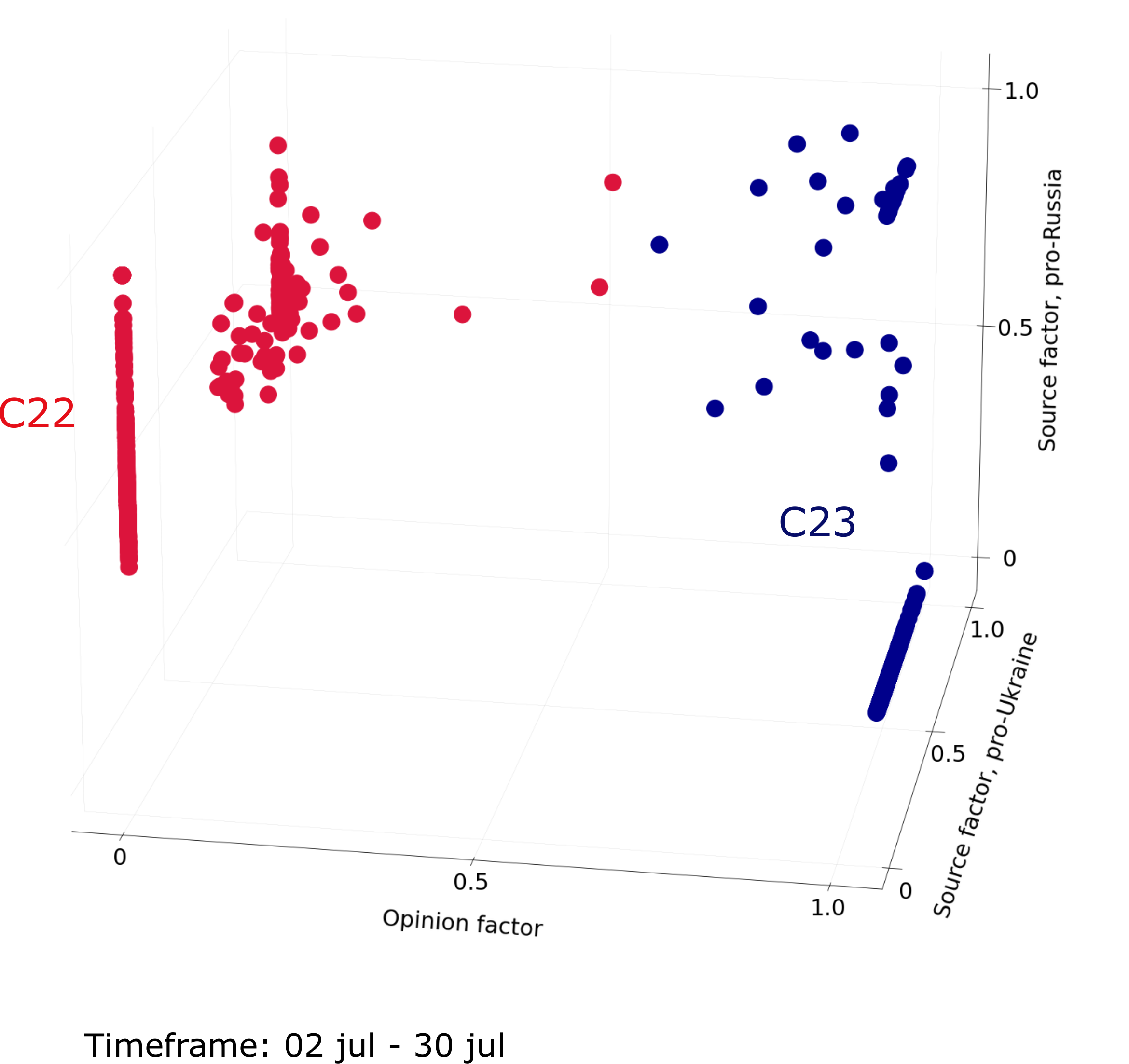}
  \caption{2 clusters}
  \label{fig:2clusters_ukraine}
\end{subfigure}
\caption{\textbf{Clusters of users interacting about the Ukraine conflict debate (n=784 users active on at least 80\% of timeframes).} Figure (a) presents the clusters identified among users during a 3-cluster timeframe, extending from February 12 to March 12, 2022. The proportions of users in each cluster are as follows: C15 = 50\%, C16 = 10\%, C17 = 40\%. Figure (b) presents the clusters identified among users during a 4-cluster timeframe, extending from March 12 to April 9, 2022. The proportions of users in each cluster are as follows: C18 = 42\%, C19 = 14\%, C20 = 4\%, C21 = 40\%. Figure (c) presents the clusters identified among users during a 2-clusters timeframe, extending from July 2 to July 30, 2022. The proportions of users in each cluster are as follows: C22 = 56\%, C23 = 44\%.}
\label{fig:clusters_ukraine}
\end{figure}

For the Ukraine conflict emerging debate, the number of clusters identified is much more variable, ranging between 2 and 9 clusters. Before the conflict was officially declared (first and second timeframes), the number of clusters is high: 8 and 9 clusters are identified. From the third timeframe, the number of clusters greatly reduces, with only three clusters differentiated (C15, C16, and C17 in Figure \ref{fig:3clusters_ukraine}). As in the case of the aggregate analysis, two clusters of polarized users are identified, but the difference lies in the clustering of intermediate users who are gathered into a unique cluster (C16). Right away, from the beginning of March, four clusters are identified and last from March to June (C18, C19, C20, and C21 in Figure \ref{fig:4clusters_ukraine}). During these consecutive timeframes, clusters are similar to those identified during the aggregate analysis (C5, C6, C7 and C8 in Figure \ref{fig:clusters_aggregate_ukraine}), with two clusters of polarized users, and two clusters of intermediate users. Finally, during the last two timeframes, only two clusters are discriminated (C22 and C23 in Figure \ref{fig:2clusters_ukraine}). As for the last timeframes of the COVID-19 vaccine debate, all users are identified as polarized users.

We can conclude that analyzing polarization as an evolving process contributes to highlight that Twitter users' polarization evolves over time, as well as the clusters they form, both in terms of number and nature.  While a population of stable polarized users is maintained, presumably comprising the most politicized individuals, one or more intermediate sets of individuals are subject to change. Besides, this time-aware analysis contributes to differentiate between debates, which were identified as similar with the aggregate analysis. While some similarities can be noticed between the two debates, their overall evolution is different. These differences could be associated with their maturity. This strengthens our conviction of considering polarization as a process, not only as a state.

\subsection{A period-based analysis of polarization}
\label{subsec:period-based}

To gain a better understanding of the polarization process,  we now focus on the structure of the previously identified changes over time. We start by defining a period as a sequence of consecutive timeframes with consistent clusters. Concretely, two sets of clusters are considered as consistent if their interpretation is the same.

In both debates, periods are actually identified. For example, in the COVID-19 vaccine debate, a first period made up of 4 clusters persists over 3 timeframes. A second period, made up of 2 clusters, remains during the following 5 months. In the Ukraine conflict debate, a greater number of periods is identified: a first 1.5 month long period is made up of numerous clusters, followed by a short 3-cluster period, and then by two periods of 4 and 2 clusters. 

In order to understand the transition from one period to another, we propose to look closer at the evolution of the distribution of users between periods. In particular, we focus on intermediate users, who are those who vary the most over time. For the COVID-19 debate, we see that intermediate users from the first period (C10 and C11 in Figure \ref{fig:4clusters_vaccine}), have moved closer to polarized users, ultimately forming only two clusters of polarized users in the last period (C13 and C14 in Figure \ref{fig:2clusters_vaccine}). For the Ukraine conflict debate, we see that intermediate users who have no preferred community in the second period (C16 in Figure \ref{fig:3clusters_ukraine}) then split into two clusters of intermediate users (C19 and C20 in Figure \ref{fig:4clusters_ukraine}), finally getting closer to polarized users and thus being identified as polarized during the last timeframes (C22 and C23 in Figure\ref{fig:2clusters_ukraine}).

\begin{figure}[h]
\centering
\begin{subfigure}[t]{0.32\textwidth}
  \includegraphics[width=1\linewidth]{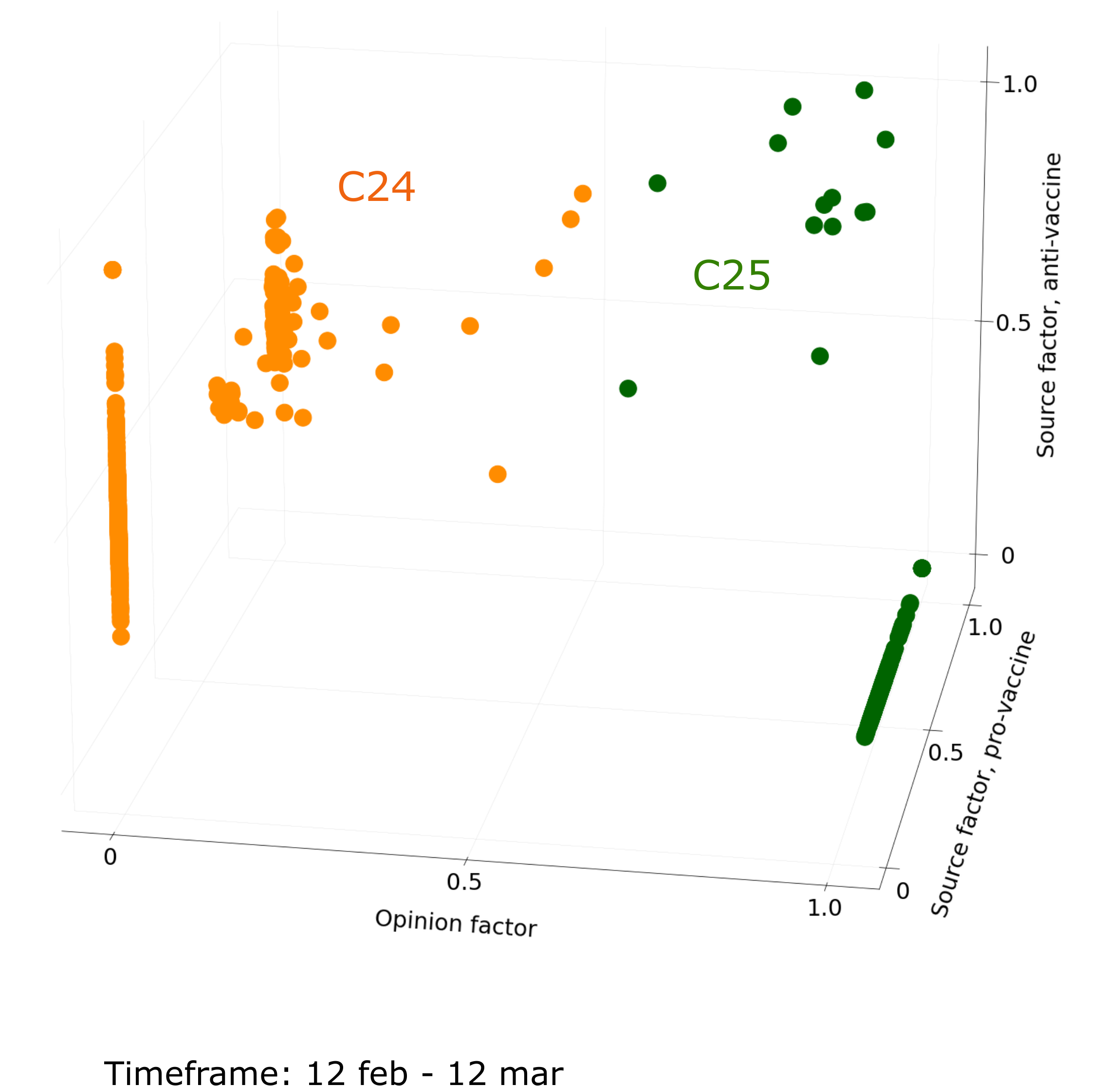}
  \caption{Initial position.}
  \label{fig:clusters_vaccine_polarized1}
\end{subfigure}
\begin{subfigure}[t]{0.32\textwidth}
  \includegraphics[width=1\linewidth]{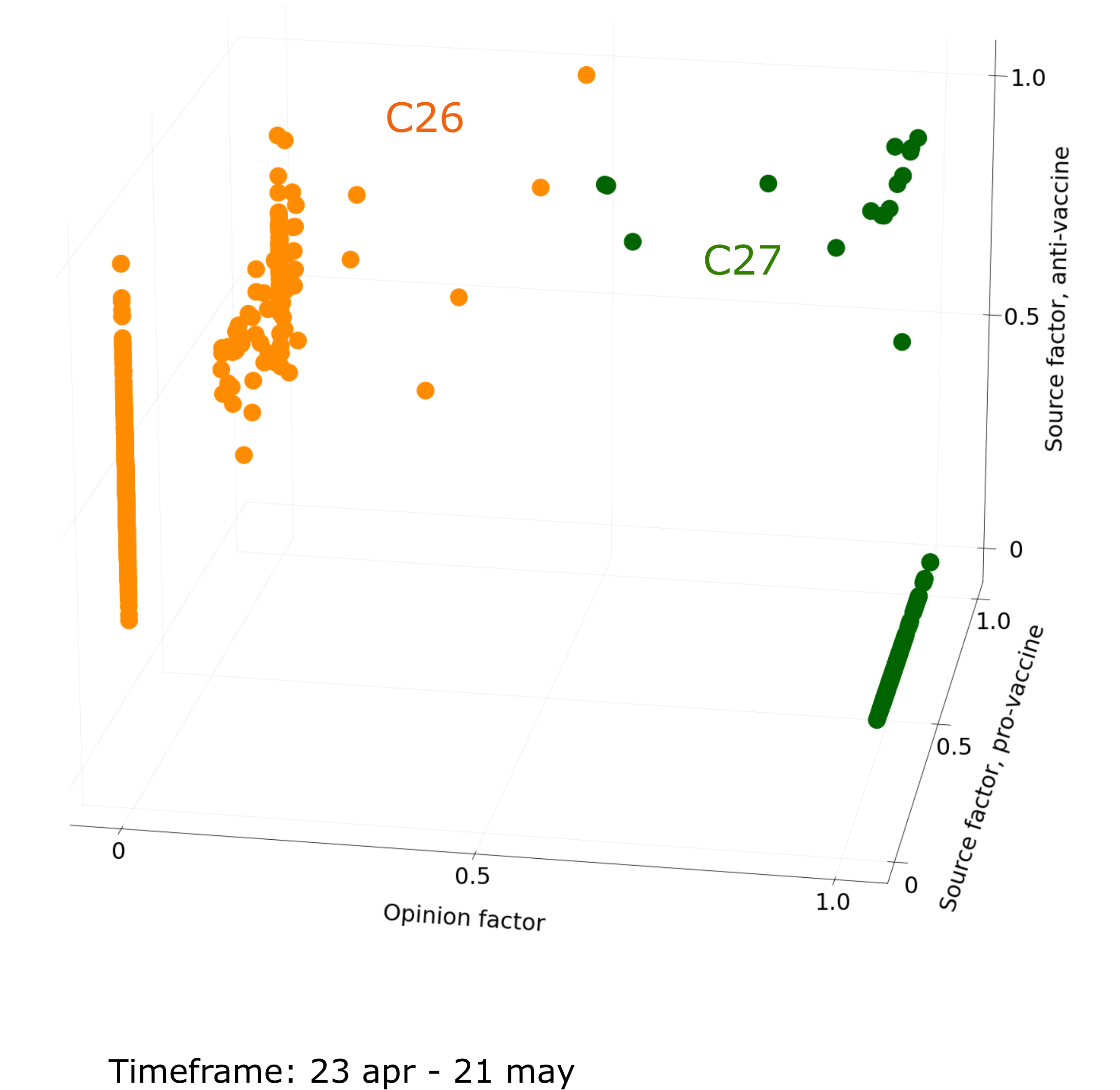}
  \caption{Intermediary position.}
  \label{fig:clusters_vaccine_polarized2}
\end{subfigure}
\begin{subfigure}[t]{0.32\textwidth}
  \includegraphics[width=1\linewidth]{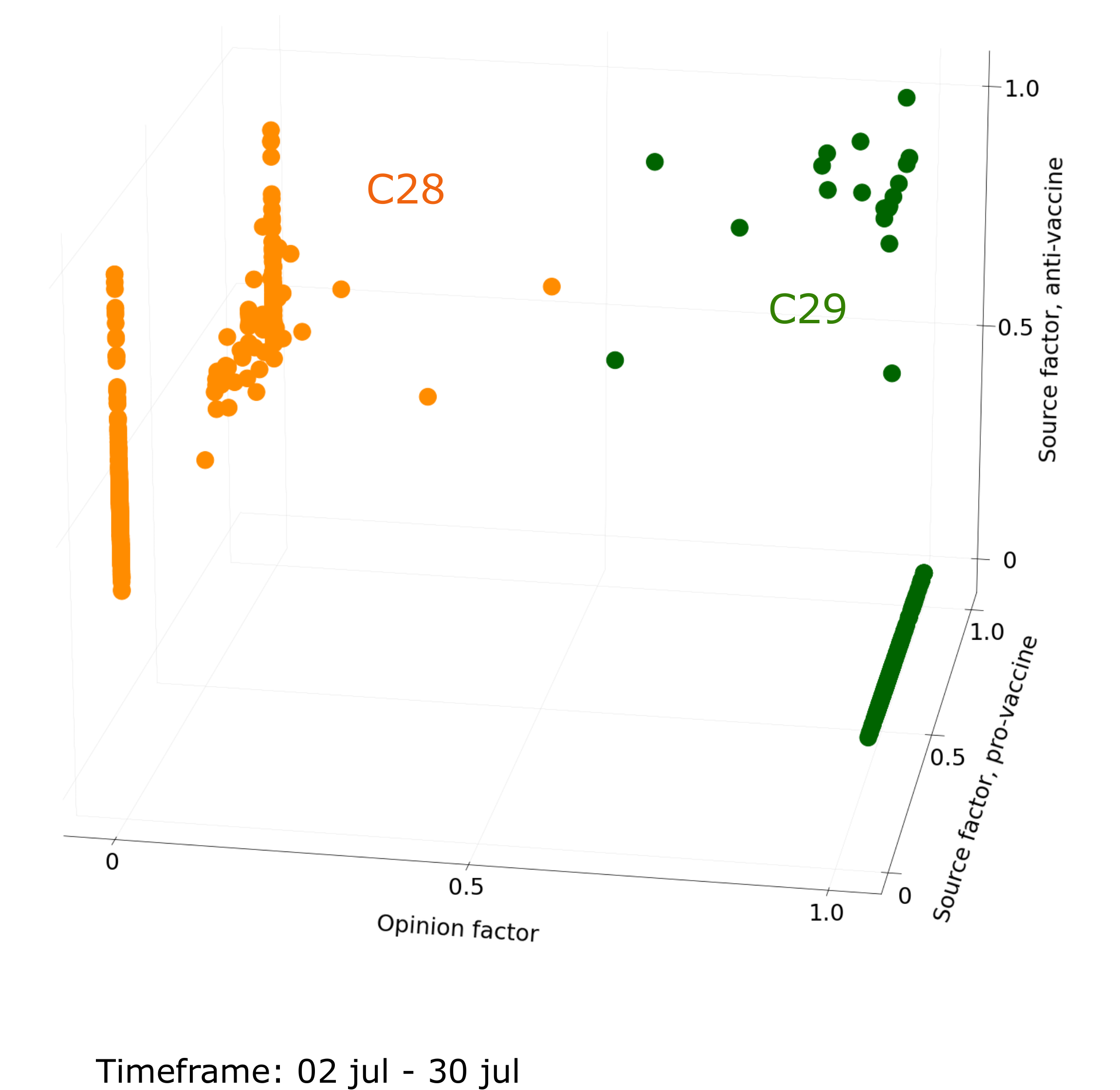}
  \caption{Final position.}
  \label{fig:clusters_vaccine_polarized3}
\end{subfigure}
\caption{\textbf{Evolution of users interacting with the COVID-19 vaccine debate during the longest period (n=685 users active on at least 80\% of timeframes).} Figure (a) shows the initial position of users during a timeframe extending from February 12 to March 12, 2022. Figure (b) shows the intermediary position of users during a timeframe extending from April 23 to May 21, 2022. Figure (c) shows the final position of users during a timeframe extending from July 2 to July 30, 2022.}
\label{fig:clusters_vaccine_polarized}
\end{figure}

\begin{figure}[h]
\centering
\begin{subfigure}[t]{0.32\textwidth}
  \includegraphics[width=1\linewidth]{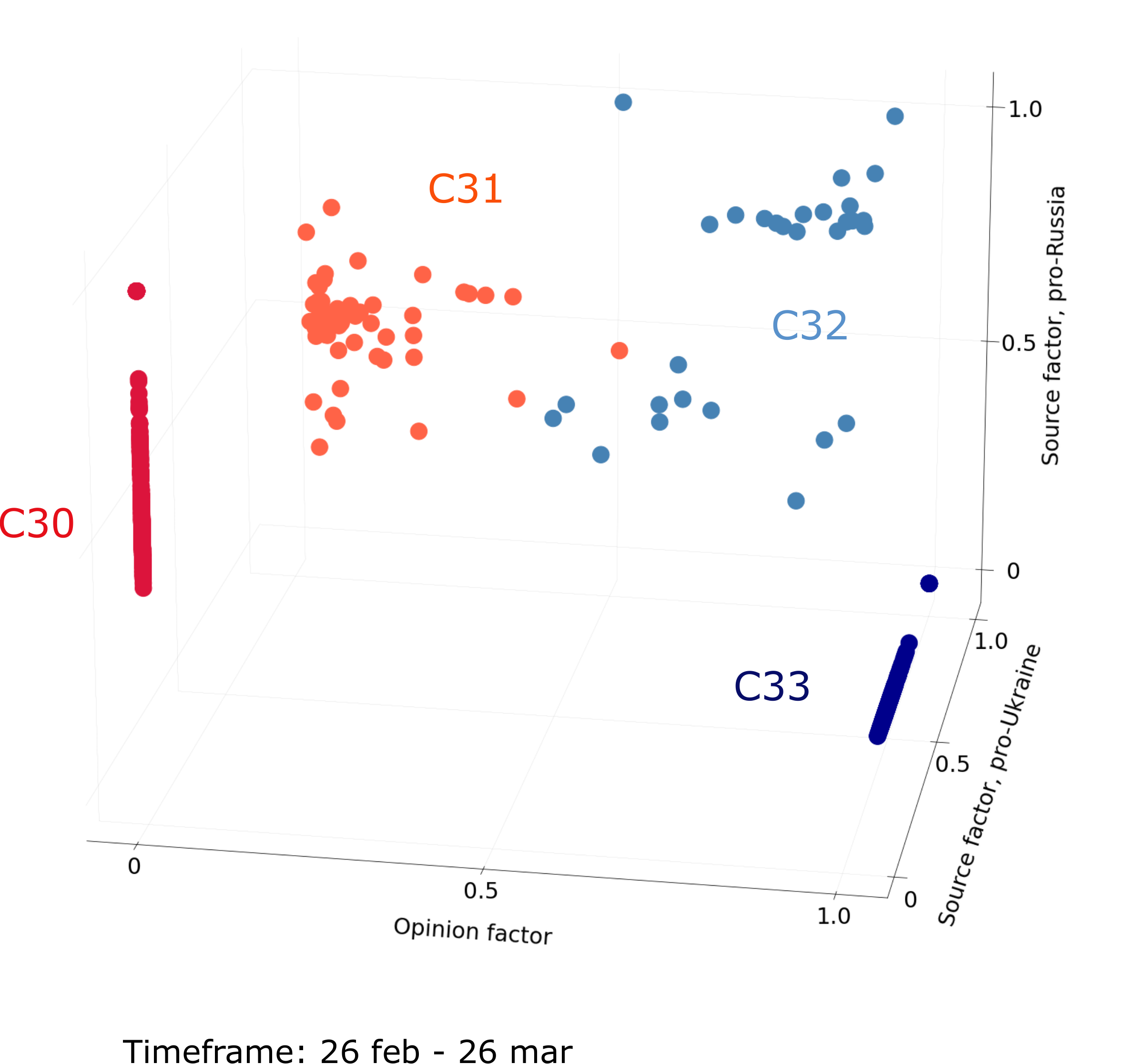}
  \caption{Initial position.}
  \label{fig:clusters_ukraine_convergence1}
\end{subfigure}
\begin{subfigure}[t]{0.32\textwidth}
  \includegraphics[width=1\linewidth]{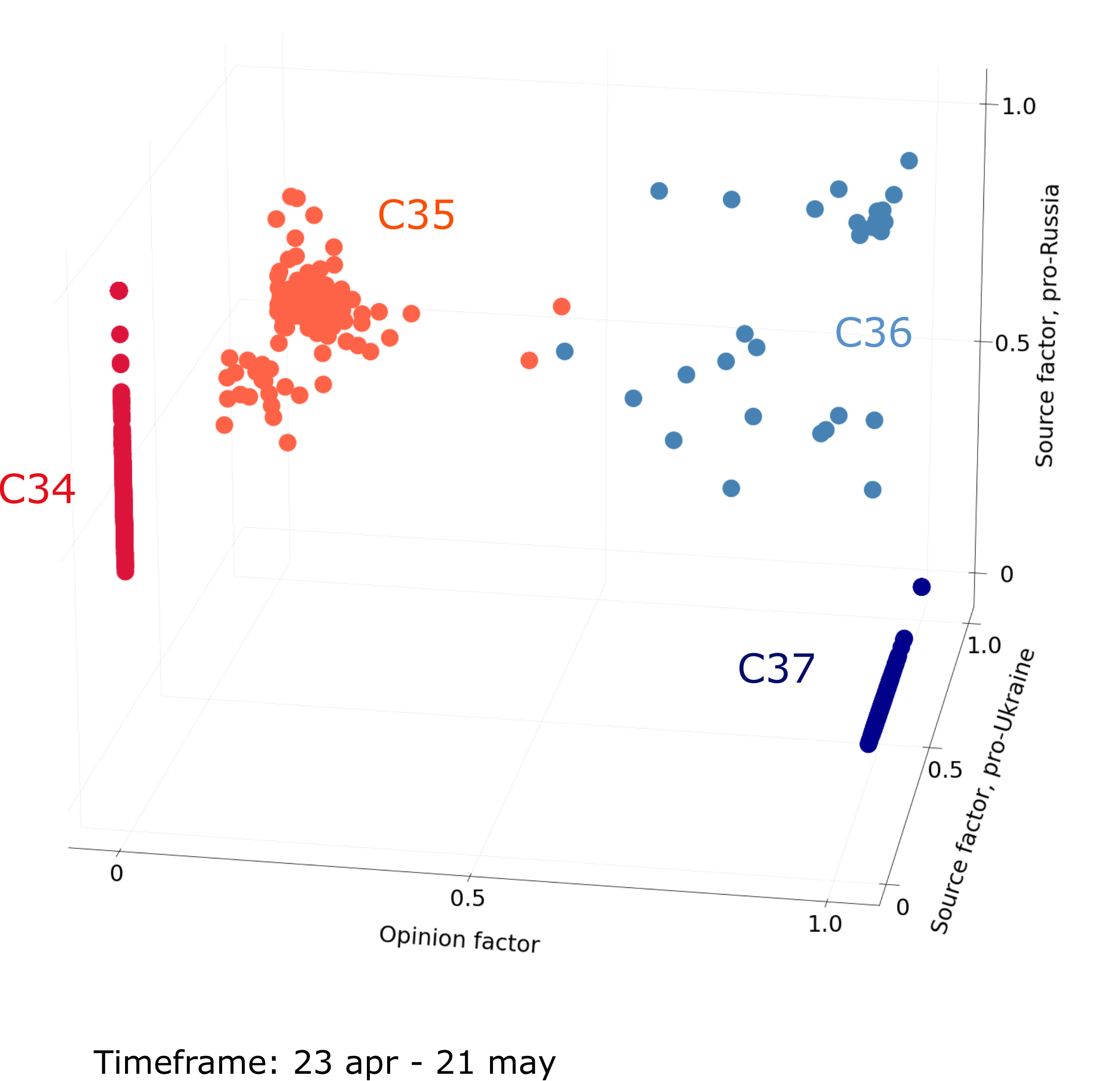}
  \caption{Intermediary position.}
  \label{fig:clusters_ukraine_convergence2}
\end{subfigure}
\begin{subfigure}[t]{0.32\textwidth}
  \includegraphics[width=1\linewidth]{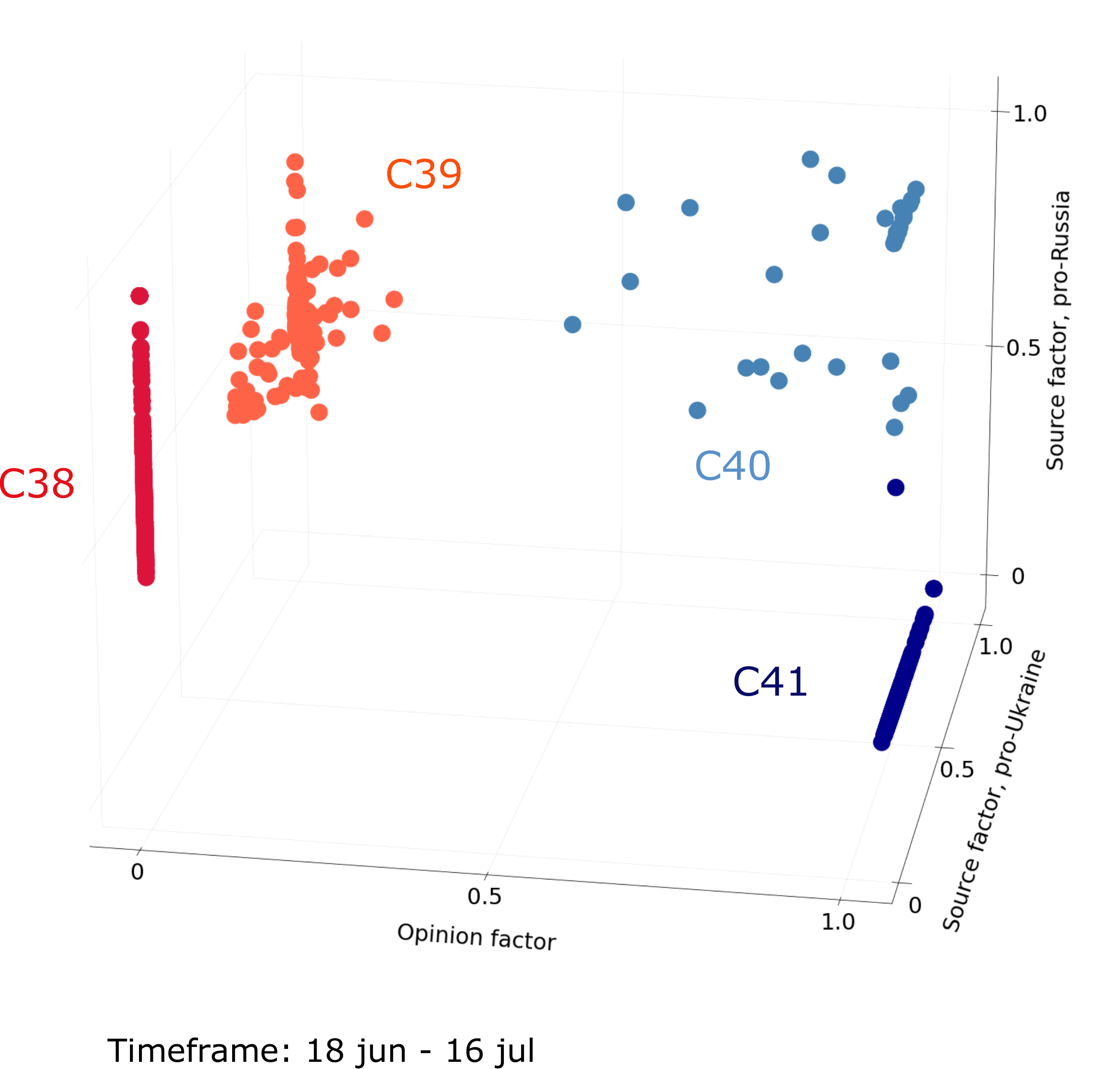}
  \caption{Final position.}
  \label{fig:clusters_ukraine_convergence3}
\end{subfigure}
\caption{\textbf{Evolution of users interacting with the Ukraine conflict debate during the longest period (n=784 users active on at least 80\% of timeframes).} Figure (a) shows the initial position of users during a timeframe extending from February 26 to March 26, 2022. Figure (b) shows the intermediary position of users during a timeframe extending from April 23 to May 21, 2022. Figure (c) shows the final position of users during a timeframe extending from June 18 to July 16, 2022.}
\label{fig:clusters_ukraine_convergence}
\end{figure}

To go further, we analyze how the users evolves within the longest periods identified for each debate. Looking first at the evolution of users during the 5 months long period for the COVID-19 vaccine debate (Figure \ref{fig:clusters_vaccine_polarized}), users continue to polarize even more over time, especially those who were identified as intermediate during the previous period. The interactions between opposing communities are thus drastically reduced, and this divergence gradually increases during this period. 
About the Ukraine conflict debate, looking at the evolution of the distribution of users during the longest period (See Figure \ref{fig:clusters_ukraine_convergence}), we confirm that intermediate users are gradually moving closer to polarized users. This reflects a drop in interactions in the community that is not their preferred community. Once the imbalance between is too important, and intermediate users are too close to polarized users, this convergence period ends and forms only two clusters of polarized users.

Here we characterize the set of previously identified periods:
\begin{itemize}
    \item \textbf{Unstructured periods} are periods with numerous clusters, among which no specific behaviors can be identified. In the Ukraine conflict debate, it corresponds to the first period. This period has no instance in the more settled COVID-19 vaccine debate. 
    \item \textbf{Balanced periods} are periods made up of three clusters: two clusters of users adopting clear-cut positions and one cluster of users maintaining balanced interactions in the two opposing communities. This period is the second one identified for the Ukraine conflict debate, while it has no instance in the period analyzed for the COVID-19 vaccine debate.
    \item \textbf{Convergence periods} are periods where four clusters are differentiated: two clusters of polarized users, and two clusters of intermediate users. The latter are getting closer to polarized users over the period. Such a period is identified in both debates. 
    \item \textbf{Polarized periods} are periods composed of only two clusters of polarized users, belonging to either community.  Such a period is identified in both debates.
\end{itemize}

If one steps back, the difference in periods between the two debates can be explained by the maturity of the debates which probably influences the way users interact with each other. In the Ukraine conflict debate, the unstructured and balanced periods are identified, which translates the clear distinction between polarization ahead of the emergence of the debate (unstructured period), and polarization from the start of the debate (balanced period). These two periods are not identified for the COVID-19 vaccine debate, which was discussed for a long time at the time of the data collection. Balanced users thus do not exist anymore in the COVID-19 vaccine debate as most of the users already have a preferred side of the debate. 

\begin{figure}[h!]
\centering
\begin{subfigure}[t]{1\textwidth}
  \includegraphics[width=1\linewidth]{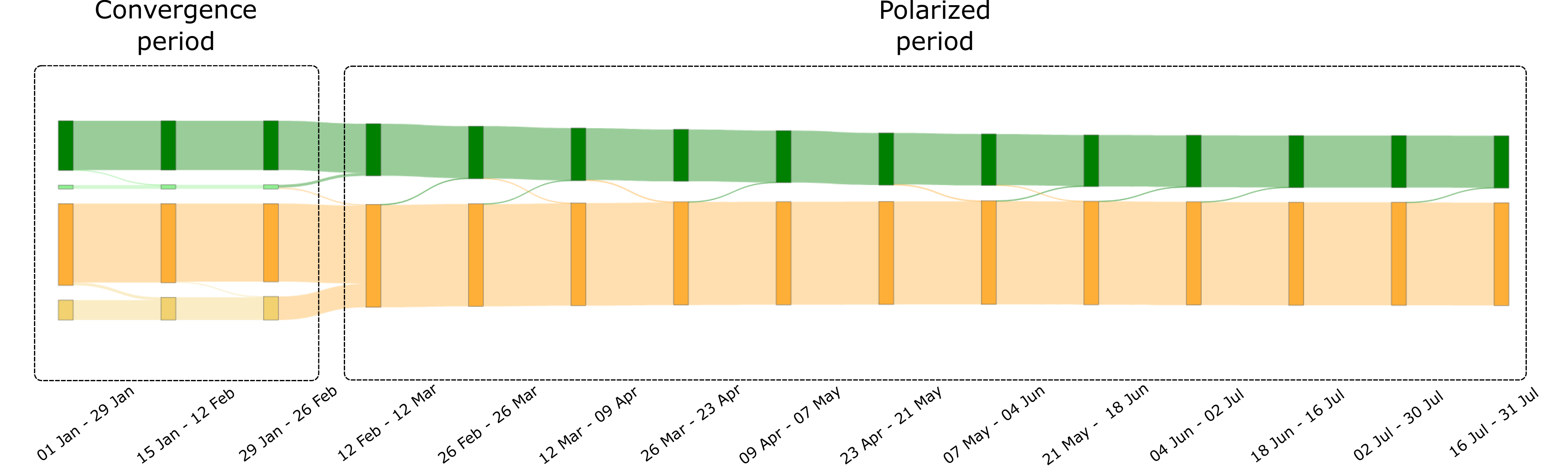}
  \caption{COVID-19 vaccine debate.}
  \label{fig:sankey_vaccine}
\end{subfigure}
\begin{subfigure}[t]{1\textwidth}
    \centering
    \includegraphics[width=1\linewidth]{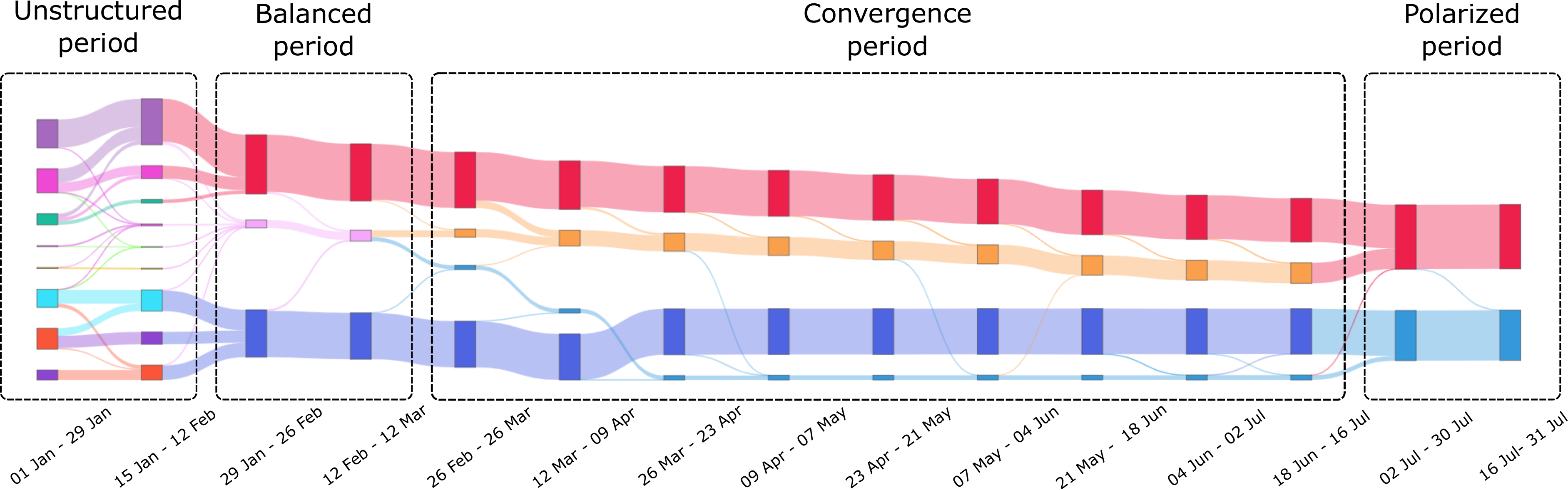}
    \caption{Ukraine conflict debate.}
    \label{fig:sankey_ukraine}
\end{subfigure}
\caption{\textbf{Sankey diagrams of the evolution of identified clusters during different periods.} Figure \textbf{(a)} presents the evolution of the number of clusters during convergence and polarized periods for the COVID-19 vaccine debate, and figure \textbf{(b)} presents the evolution of the number of clusters during the unstructured, disturbance, convergence and polarized periods for the Ukraine conflict debate.}
\label{fig:sankey}
\end{figure}

Nevertheless, a specific pattern of periods is shared by both debates: the convergence period is systematically followed by the polarized period. This pattern reflects a gradual decline in the interest of intermediate users in the community that is not their major community, and users systematically end up polarized. Intermediate users therefore do not appear to be stable over time. However, the duration of the periods in this pattern varies between debates. Here again, the maturity of the debate can explain the difference. The sequence of periods and associated clusters are presented in Figure \ref{fig:sankey}. 

To summarize, this period-based analysis contributes to highlight clear similarities between the two debates. Yet, we have highlighted the existence of specific periods, some of them forming a pattern common to both debates. This makes us think user polarization follows a predefined pattern, independent of the debate and its maturity. To go further, an analysis of the contexts underlying the temporal evolution of clusters could help to better characterize polarization dynamics.

\subsection{A context-aware analysis of polarization}
\label{subsec:context_aware}

User behavior obviously takes place within a broader context. We hypothesize that the differences previously outlined between debates are in part the result of external or debate-related events and we discuss this hypothesis here.  

We would like to first get back to the unstructured period identified for the Ukraine conflict debate only.  Even if the tensions between Russia and Ukraine were discussed before February 2022, the media framing of this topic was secondary at the time, and did not promote opinion crystallization. It is thus probable that people were not much aware of the debate.  As soon as the war was declared, this unstructured period ended, followed by the balanced period. The balanced position of intermediate users in this period is transient, and results in the emergence of imbalanced interactions between the two communities in an associated extended convergence period. This long period can be explained by the fact that news media have devoted considerable space to covering this topic in Spring 2022. This stimulates discussion with close friends and family, keeps users aware of the debate, but it does not necessarily change the ideas already shared \cite{lazarsfeld1968people}.

About the COVID-19 vaccine debate, the lack of unstructured and balanced periods does not imply that they did not occur, they may simply have occurred prior to the period of study. For the specific case of the unstructured period, it is probable that it did not appear some time before January, 2022. Indeed, the discussions about the vaccine debate reemerged with the pandemic, and the debate was already active.
The identification of a convergence period is quite surprising, as it was expected that the debate would be stable at the beginning of 2022. Looking at the French events in these times, it turns out that a measure was taken some time before the starting date of the dataset. In fact, the vaccination campaign was opened to children aged 5 to 11 on December 21, 2021. This was covered by the media and thus sparked discussions on Twitter, making some users more aware of discussions from their opposite community. The question of the possible existence of a balanced period at the end of December remains. In all cases, the convergence period has started at the latest on December 21, it is thus at most 1.5 months long. It makes a significant difference with the duration of this period in the Ukraine conflict debate (5 months).
From a general perspective, we can conclude that in each debate the emergence of intermediate users (balanced or convergence periods) is triggered by a disruptive event. 
On one side, the invasion of Ukraine by Russia, which is a sudden event in an immature debate. It has led to mass mobilization, had a major impact on society, and generated massive reactions. On the other side, a governmental decision linked to the COVID-19 vaccine, which is not sudden (an announcement has been made some days before its implementation) and occurs in a well-established debate. These events are supported by a modified media framing, and an increase in Twitter users’ activity on the related debate. However, the impact of both events differs: in the emerging debate, it first leads to a balanced period, where a significant proportion of users show no preference for either community, followed by a convergence period lasting several months. In the mature debate it encourages some users, probably polarized, to seek information in the opposing community during a short period of convergence. In our view, the maturity of the debate and the suddenness of the event are the two main reasons for this difference.
It confirms that users with a constructed opinion on a societal debate can remain aware of the evolution of the debate, by getting interested in the opposing community. However, the more crystallized this opinion, the faster they go back to their community. 

This context-aware analysis has shown that, whatever the maturity of the topic, some events may impact user polarization, the duration of which depends on the maturity of the debate and the type of event. We can thus suppose that debates, even when opinions are crystallized, may cyclically have intermediate users appearing, triggered by external events. 

\section{Discussion}
\label{sec:discussion}
The analyses conducted in this work aimed to deconstruct the notion of polarization and refine the distinction between affective and ideological polarization. Within a population of Twitter users that tends to be more politicized than the average, the identification of several groups of users with specific polarization behavior, provides the potential for the distinction of different crystallized states of polarization. This crystallization is not only expressed by very decided opinions on specific subjects, but by a tendency to reduce the information spectrum to an echo chamber or to aggressively refuse all contradiction. It is known to have a major impact on the reliability of opinion surveys, and seems to go through different phases before stabilizing. The temporal analysis contributes to distinguish between "stable” and “unstable” polarized users. We assume that the former are the most ideologically polarized. In this sense, we use the term "politicized". The latter, qualified as intermediaries, are individuals likely to be affectively polarized. But this may be the result of a transient state linked either to the maturity of the debate, or to the introduction of an event that forces users to reposition themselves.
Our work allows an unprecedented identification of periods that contributes to deconstruct polarization: polarization is not a simple state, it is an evolving process. In fact, this work also contributes to better measure the way context-related events can modify the framing of a debate and provoke a systematic dispersion of a subset of Twitter users (intermediate users, affectively polarized) before they gradually polarize, whatever the nature of the debate (mature or emerging). We do not rule out the possibility that this repositioning can be due to the specificity of the Twitter population, \textit{i.e.} a population particularly sensitive to the effects of media framing. However, our findings extend those from Waller and Anderson \cite{waller2021quantifying}, that highlight an alignment between online polarization and external events in the US context, to the French-context.

The nature and time of broadcast of the information provided in the debate is therefore crucial, and can have a different effect on users: undifferentiated when the debate is recent (a single intermediate cluster), more targeted when the debate is long-standing (two intermediate clusters). We assume that the more the debate is fueled by elements of consensus, the less affective polarization has a hold, and therefore the more the media framing (that is particularly structuring on a Twitter population) has to be argued to facilitate an ideological polarization likely to fuel a politicization process (reducing the number of intermediary groups). The more sudden, unexpected and unpredictable changes in frameworks, the more destabilized users, who take time to (re)position themselves, even when the debate is mature (COVID). Our work leads to similar conclusions that can be read through Goffmanian “secondary frames”\cite{goffman1974frame}, so media frames can overlap, leading to periods of hesitation, confrontation, and repositioning. 

We wonder how we can take advantage of this unprecedented understanding and characterization of the polarization process to reduce animosity between opposing communities, while keeping users properly informed. The aim is obviously not to manipulate the opinions of social media users, but to keep them informed as widely as possible, so that they can make informed decisions in a democratic context. Recall that the literature used to address the problem by diversifying news recommendations. This single strategy does not have the same impact on all users and can even reinforce polarization \cite{bail2018opposing,treuillier2022being}. The findings from this work are an opportunity to rethink diversification through multiple strategies. Concretely, the convergence and balanced periods, in which intermediate users appear, are probably the periods during which personalized and diverse recommendations can be proposed by adopting recommendation strategies that differ between intermediate and polarized users. We can for example imagine confronting polarized users with other debates, \textit{i.e.} providing content diversity, so as to draw their attention to something different and avoid further polarization. In addition, recommendations could be adapted to the temporality of the debate, \textit{i.e.} acting at the most appropriate moment to maintain intermediates with diversified sources of information, thus helping to foster a healthy debate and reduce animosity. Finally, the appearance of a new debate is an opportunity to promote diversity of opinions so that users can quickly understand all the positions expressed on the subject.

Looking ahead, we expect to compare the debates studied with debates offering a clearer political reading. Indeed, both debates - the COVID vaccine and the  Ukraine conflict - have been subject to considerable disruption. For the former, the spread of conspiracy theories, and for the latter, strong anti-Americanism, have largely blurred the traditional political cleavages between left and right. It would therefore be interesting to carry out the analysis again by choosing more politically divisive topics (taxation, interventionism versus liberalism, etc.) to better identify periods of repositioning and convergence. It also appears that this temporal analysis will benefit from the study on periods of greater crystallization of opinions, such as election campaigns, which are much more sensitive to the effects of media framing, and over a longer time.

\section{Methods}
\label{sec:methods}

\subsection{Data}
\label{subsec:data}

We used the Twitter API (\textit{v2}), with academic research access to collect data. Our methodology relies on the concept of elite users~\cite{primario2017measuring} that represent users who are relevant to the subject matter. We assume that elite users' tweets are in line with their beliefs about the selected debate. Inspired by the methodology of Primario~\textit{et al.}~\cite{primario2017measuring}, we fix conditions to select legitimate elite users: they need to (1) have a significant number of followers; (2) personally manage their Twitter account; (3) are known by the general audience, through media or government interventions; and (4) are qualified by education and/or profession to address the subject of matter.  

Elite users are an effective entry point for collecting data about a specific topic because their opinions are publicly known~\cite{primario2017measuring}. Nevertheless, our objective is to analyze the interaction behaviors of standard users (non-elite). It is thus necessary to have a dataset that is balanced in terms of opinion carriers, but also representative of standard users' behaviors on social media about a specific debate and during a specific time span. To build such a quality dataset, we carried out several steps, run after having chosen the debates, identified a relevant set of elite users, and defined a collection time span. These steps are as follows: (1) Collect all tweets published by elite users during the predefined period; (2) Filter tweets about the topic of interest; (3) Collect information about a random subset of interacting standard users for each collected tweet; (4) Identify the most active standard users among those selected in Step~3; (5) Collect all interactions of selected standard users on collected elite users' tweets during the defined period; (6) Among all collected interactions, filter those that are related to the tweets collected in step (1).

We collected data about two topics: the COVID-19 vaccine and the Ukraine conflict. Following the procedure detailed above, we manually identified 20 French-speaking elite users having a legitimate voice in the vaccine debate (10 pro-vaccine and 10 anti-vaccine), and 20 other French-speaking users expressing themselves about the Ukraine conflict (10 pro-Ukraine and 10 pro-Russia). Their opinion is known because they have clearly expressed it publicly, and the community to which they relate is therefore unambiguous. To preserve their confidentiality and meet Twitter policy, we do not share the names or usernames of the selected accounts. We collected all elite users' tweets over a 7-month time span, extending from January 1, 2022, to July 31, 2022. 

Based on relevant debate-related French hashtags, either for the COVID-19 vaccine or the Ukraine conflict debate, and a random tweet corpus~\cite{turenne2018rumour}, we trained a two-class classifier based on BertTweetFR~\cite{guo-etal-2021-bertweetfr}. This classifier allowed us to keep only elite users' tweets dealing with the selected debates. Here, we focus on retweets, which are signs of approval and thus give information about what users agree with~\cite{conover2011political}. 

Following this methodology, we collected information about 100 randomly selected retweeters for each collected tweet, which we hope to be representative of all users. Among the selected retweeters, we focused on the 1,000 most active ones in each debate (500 pro-vaccine / 500 anti-vaccine, 500 pro-Ukraine / 500 pro-Russia). All in all, the collected dataset contains 299,879 retweets about the COVID-19 vaccine debate (16,791 retweets in the pro-vaccine community, 283,088 in the anti-vaccine community), and 152,802 about the Ukraine conflict debate (41,631 retweets in the pro-Ukraine community, 111,171 in the pro-Russia community), made by a set of 1,000 standard users on 20 elite users' tweets for each debate.

\subsection{Evaluation of polarization on social media}
\label{subsec:polarization_evaluation}

We rely on three factors to measure polarization. First, we study the opinions that are shared by standard users, according to the communities within which they interact. As we study two bi-community debates, we secondly study the diversity of sources with which they interact in one community, and the diversity of sources with which they interact in the confronting community. We can represent each factor as a probability distribution, specific to each user, and then compute normalized entropy $H_N$:
\begin{equation}
\label{eq:entropy}
  H_N(X)= \frac{-\sum_{x} P(x)log(P(x))}{log(n)}
\end{equation}

Where $X$ is a discrete random variable that takes $n$ possible values, and $P(x)$ is the probability of entity $x$. To ensure that higher computed values are associated with high polarization, we use $H' = 1-H_N$. 

To make the opinion factor gives an indication about the community a user belongs, the opinion factor is oriented \cite{treuillier23}. To this end, the sign of the normalized entropy computed based on the probability distribution of interactions within both considered communities is set according to the community within which a user interacts more. We fix positive values for polarization in the pro-vaccine or pro-Ukraine communities, and negative values for polarization in the anti-vaccine or pro-Russia communities. Finally, we then apply the following transformation so that the values lie within $[0,1]$:

\begin{equation}
    H'= \frac{H^\pm+1}{2}
\end{equation}

This way, a $H'$ value close to $0$ represents a user  either closer to the anti-vaccine or pro-Russia community depending on the considered debate, while $H' = 1$ indicates a user  closer to the pro-vaccine or pro-Ukraine community. Besides, as the source factors are specific to each community, they are not oriented, and range in $[0,1]$. A source factor equal to $0$ indicates that the user interacts only with one source in the community. The more she interacts with diverse sources in that community, the higher the source factor. 

For the aggregate analysis of polarization, these factors are computed for each user during the entire timespan, between January 2022 and July 2022. For the time-aware analyses of polarization, factors are computed over specific timeframes. 

\subsection{Definition of timeframes for the time-aware analysis}
\label{subsec:time_frames}

To study the evolution of polarization behaviors, we defined 4-week sliding timeframes, with a 2-week overlap. Over the 7 months of data collection,  15 timeframes are thus formed between January 1 and July 31, 2022. Polarization factors, about opinions and sources, can thus be measured on every successive timeframe.

Of course, some users are not active on all defined timeframes. We have defined a threshold of 20\% inactive periods, and users who are inactive more than 20\% of the timeframes are removed from the dataset in the time-aware analysis. Among the initial 1,000 users studied in the aggregate analysis, we thus kept 685 users for the COVID-19 vaccine debate, and 784 users for the Ukraine conflict debate. 

\subsection{Identifying polarization behavioral classes}
\label{subsec:kmeans}

To discriminate between polarization behavioral classes, we use the $k$-means~\cite{likas_global_2003} algorithm. The number of clusters $k$ is optimized by maximizing 2 traditional metrics: Davies-Bouldin Index~\cite{davies1979cluster} (the lower the better) and Silhouette Index~\cite{rousseeuw_silhouettes_1987} (the higher the better). The clustering algorithm is applied to the three factors of polarization computed for each user (opinion factor and both sources factors).

However, as we study polarizing debates, some users may remain tightly bunched for some of the factors. This limits the differentiation between them. In order to improve clustering performance, we apply a polynomial transformation having a sigmoid-like pattern to the three factors, as follows:
\begin{equation}
\label{polynomial transformation}
    f(x) = \frac{x^a}{x^a + (1-x)^a}
\end{equation}
The parameter $a$ allows to control the stiffness of the curve and to better discriminate either extreme values or intermediate values. Besides, we assumed that each factor can have different weightings in the evaluation of polarization. This means that factors can have different weights in the computation of the Euclidean distance on which the $k$-means algorithm relies.
Thus, both the parameter $a$ of equation \ref{polynomial transformation} and the weights of the three factors were optimized through the clustering process. We selected values that allowed higher clustering performances. 
The optimal value of $a$ is  $a = 0.5$ for the COVID-19 vaccine debate, while $a = 0.33$ for the Ukraine conflict debate.  For both debates, the source factor has a weight of $60\%$, while the $40\%$ remaining is distributed among the two source factors. 

We use the same algorithm and parameters to discriminate between behavioral classes over each timeframe for the time-aware analysis.

\paragraph{Data and Code availability}
The source code of the study is available at the following public GitHub repository: \url{https://github.com/Celina-07/polarization_social_media}. 
All the analysis were performed using Python\cite{van1995python} v3.9.7, and following libraries: matplotlib\cite{Hunter:2007} v3.5.3, numpy \cite{harris2020array} v1.22.4, pandas \cite{mckinney2011pandas} v1.5.3, plotly v5.4.0,pydlc \cite{moritz2018visualizing} v0.2., scikit-learn \cite{scikit-learn} v1.0.2, scipy \cite{2020SciPy-NMeth} v1.7.3, statsmodels \cite{seabold2010statsmodels} v0.13.5,  tqdm \cite{da2019tqdm} v4.62.3.

Minimal datasets required to replicate the methods presented in this paper are available on the GitHub repository. Due to the terms of data license agreement signed with Twitter, Inc., all data are not publicly available. Nevertheless, all relevant data are available upon reasonable request to the authors. 

\paragraph{Aknowledgments}
This research was supported by BOOM (Modeling and Opening Opinion Bubbles) (ANR-20-CE23-0024).  

\bibliographystyle{unsrt}
\bibliography{refs}

\end{document}